%% file: axialsep.tex
\documentclass[prc,twocolumn,showpacs,10pt,floatfix]{revtex4}

\usepackage{graphicx}
\usepackage{amsmath}
\usepackage{amssymb}
\usepackage{citesort}
\usepackage{mathrsfs}
\providecommand{\mathscr}{\mathcal}  

\usepackage{closedcases}

\newcommand{\<}[1]{\hspace{-0.11111em}#1\hspace{-0.11111em}}
\newcommand{\rtrim}[1]{#1\hspace{-0.11111em}}

\DeclareRobustCommand{\grp}[1]{\mathrm{#1}}
\DeclareRobustCommand{\efive}{\grp{E}(5)}
\DeclareRobustCommand{\xfive}{\grp{X}(5)}

\DeclareRobustCommand{\grpsuthree}{\grp{SU}(3)}

\begin{document}


\title{
Effects of $\beta$-$\gamma$ coupling in transitional nuclei
and the validity of the approximate separation of variables}

\author{M. A. Caprio}
\affiliation{Center for Theoretical Physics, Sloane Physics Laboratory, 
Yale University, New Haven, Connecticut 06520-8120, USA}

\date{\today}

\begin{abstract}
Exact numerical diagonalization is carried out for the Bohr
Hamiltonian with a $\beta$-soft, axially stabilized potential.
Wave function and observable properties are found to be dominated by
strong $\beta$-$\gamma$ coupling effects.  The validity of the
approximate separation of variables introduced with the $\xfive$
model, extensively applied in recent analyses of axially stabilized
transitional nuclei, is examined, and the reasons for its breakdown
are analyzed.
\end{abstract}


\maketitle




\section{Introduction}
\label{secintro}

The Bohr Hamiltonian~\cite{bohr1998:v2} with a $\beta$-soft but
$\gamma$-stabilized potential has served as the basis of recent
investigations of the collective structure of transitional nuclei
intermediate between spherical and axially symmetric deformed
shape~\cite{iachello2001:x5}.  The solutions obtained so far for the
$\xfive$ model~\cite{iachello2001:x5,bijker2003:x5-gamma} and its
various
extensions~\cite{bonatsos2004:x5-monomial,caprio2004:swell,bonatsos2004:z5-triax,fortunato2004:beta-soft-triax,pietralla2004:confined-beta-soft,bonatsos2005:octupole-aqoa-actinides}
have relied upon an \textit{approximate} separation of variables,
introduced in Ref.~\cite{iachello2001:x5}, since solution of the exact
problem was not possible.  These approximate calculations have been
extensively compared with experimental
data~\cite{casten2001:152sm-x5,bizzeti2002:104mo-x5,kruecken2002:150nd-rdm,caprio2002:156dy-beta,dewald2002:150nd-rdm,caprio2003:diss,hutter2003:104mo106mo-rdm,clark2003:x5-search,tonev2004:154gd-rdm,mccutchan2004:162yb-beta,mccutchan2005:166hf-beta}.

However, numerical methods recently developed by Rowe \textit{et
al.}~\cite{rowe2004:tractable-collective,rowe2004:spherical-harmonics,rowe2005:collective-algebraic}
make exact numerical diagonalization of the Bohr Hamiltonian feasible
for transitional and deformed situations, without recourse to the
approximations of Ref.~\cite{iachello2001:x5}.  The solution process
involves diagonalization in a basis constructed from products of
optimally chosen $\beta$ wave functions with five-dimensional
spherical harmonics.  This method yields much more rapid convergence
in the presence of significant $\beta$ deformation than is obtained
with conventional methods~\cite{gneuss1971:gcm,eisenberg1987:v1}.

In the present work, an exact numerical solution for the $\xfive$
Hamiltonian is obtained (Sec.~\ref{secsoln}).  The results for
wave functions and observables are examined, and the properties of
$\beta$-soft nuclei are found to be dominated by a strong coupling
between the $\beta$ and $\gamma$ degrees of freedom
(Sec.~\ref{secresults}).  Since extensive prior work has been
carried out using the approximate separation of variables of
Ref.~\cite{iachello2001:x5}, this approximation is reviewed and the
reasons for its breakdown are analyzed (Sec.~\ref{secapprox}).

\section{Hamiltonian and solution method}
\label{secsoln}
\begin{figure}
\begin{center}
\includegraphics[width=0.75\hsize]{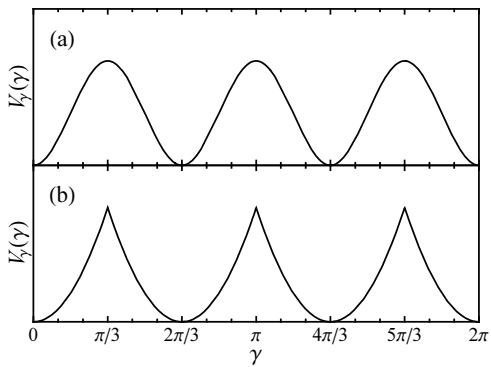}
\end{center}
\vspace{-12pt}
\caption
{ Simple forms of the potential $V_\gamma(\gamma)$ satisfying the
coordinate symmetry constraints of Ref.~\cite{eisenberg1987:v1}:
(a)~$V_\gamma(\gamma)\<=\chi(1-\cos3\gamma)$~\cite{rowe2004:tractable-collective}
and (b)~$V_\gamma(\gamma)\<=A\gamma^2$ ($0\<\leq\gamma\<\leq\pi/3$), used here for consistency with
Ref.~\cite{iachello2001:x5}. }
\label{figsawtooth}
\end{figure}

The Bohr Hamiltonian, in terms of
the quadrupole deformation variables $\beta$ and $\gamma$ and the
Euler angles $\vartheta$, is~\cite{bohr1998:v2} 
\begin{widetext}
\begin{equation}
H=- \frac{\hbar^2}{2B} \Biggl[
\frac{1}{\beta^4}
\frac{\partial}{\partial \beta}
\beta^4  \frac{\partial}{\partial \beta}
+
\frac{1}{\beta^2}
\left(
\frac{1}{\sin 3\gamma} 
\frac{\partial}{\partial \gamma} \sin 3\gamma \frac{\partial}{\partial \gamma}
 - \frac{1}{4}
\sum_\kappa \frac{\hat{L}_\kappa^{\prime2}}{\sin^2(\gamma -
\frac{2}{3} \pi \kappa
)}
\right)
\Biggr]
+V(\beta,\gamma),
\label{eqnH}
\end{equation}
\end{widetext}
where $\hat{L}_\kappa^\prime$ are the intrinsic frame angular momentum
components.  For the transition between spherical and axially
symmetric deformed structure, it is appropriate to consider a
potential $V(\beta,\gamma)$ which is soft with respect to $\beta$ 
but which provides confinement about $\gamma\<=0$.  In the $\xfive$
model, a schematic form $V(\beta,\gamma)\<=V_\beta(\beta)
+V_\gamma(\gamma)$ is used, where $V_\beta$ is taken to be a square
well potential [$V_\beta(\beta)\<=0$ for $\beta\<\leq\beta_w$ and
$\infty$ otherwise] and $V_\gamma$ provides stabilization around
$\gamma\<=0$.

Under the approximate separation of variables of
Ref.~\cite{iachello2001:x5}, most results are independent of the
specific choice of $V_\gamma$. Consequently, in
Ref.~\cite{iachello2001:x5}, the confining potential
$V_\gamma(\gamma)$ is simply described as $\rtrim\propto\gamma^2$ for
small $\gamma$.  For solution of the full problem, $V_\gamma(\gamma)$
must be defined more completely.  In general for the Bohr Hamiltonian,
the potential energy $V(\beta,\gamma)$ must be periodic in $\gamma$,
with period $2\pi/3$, and reflection symmetric about $\pi/3$, to
ensure that the potential energy is invariant under relabeling of the
intrinsic axes~\cite{eisenberg1987:v1}.  The natural choice of such
potential is $V_\gamma(\gamma)\propto(1-\cos 3\gamma)$
[Fig.~\ref{figsawtooth}(a)], as considered in
Ref.~\cite{rowe2004:tractable-collective}.  However, for consistency
with Ref.~\cite{iachello2001:x5}, in the present work the $\gamma$
potential is chosen to be the oscillator potential, given by
$V_\gamma(\gamma)\<=A\gamma^2$ on the interval
$0\<\leq\gamma\<\leq\pi/3$ and obtained outside this interval from the
symmetry requirements on $\gamma$ [Fig.~\ref{figsawtooth}(b)].

As usual for eigenproblems involving the Bohr Hamiltonian, the
parameter dependence of the solution can be simplified by an
appropriate choice of dimensionless parameters.  Transformation of the
potential as $V(\beta,\gamma)\<\rightarrow c^2V(c\beta,\gamma)$ or
multiplication of the Hamiltonian by a constant factor both leave the
solution invariant, to within an overall scale factor on the
eigenvalues and an overall dilation of all wave functions
(\textit{e.g.}, Ref.~\cite{caprio2003:gcm}).  Consequently, all
energy ratios and transition matrix elements obtained from 
diagonalization of the Hamiltonian~(\ref{eqnH}) depend only upon the
parameter combination
\begin{equation}
a\equiv\frac{2AB \beta_w^2}{\hbar^2},
\end{equation}
which measures the ``$\gamma$-stiffness'' of the Hamiltonian.  In the
remainder of this article, the notation is simplfied by setting
$\hbar^2/(2B)\<=1$ and $\beta_w\<=1$, so that $a\<=A$.  (Results for
any other values of these parameters may then be obtained according to
simple scaling relations~\cite{caprio2003:gcm}.)

For $a\<=0$, the potential is $\gamma$-independent, and the
Hamiltonian~(\ref{eqnH}) reduces to the $\efive$
Hamiltonian~\cite{iachello2000:e5}.  In this case, an exact separation
of variables occurs~\cite{wilets1956:oscillations}, and the
eigenproblem can be solved analytically~\cite{iachello2000:e5}.
Nonzero values of $a$ yield confinement around $\gamma\<=0$.  The
realistic range of values for $a$ is discussed in
Sec.~\ref{secresults}.

Rowe \textit{et
al.}~\cite{rowe2004:tractable-collective,rowe2004:spherical-harmonics,rowe2005:collective-algebraic}
have recently proposed a method for numerical diagonalization of the
Bohr Hamiltonian, with respect to an optimized product basis
constructed from the five-dimensional spherical
harmonics~\cite{bes1959:gamma,santopinto1996:so-brackets}.  Much as a
basis for solution of the Schr\"odinger problem in three dimensions
can be formed from the products of a complete set of radial functions
$f_i(r)$ with the three-dimensional spherical harmonics
$Y^L_M(\theta,\varphi)$, a basis for solution of the Bohr problem can
be constructed from products of radial basis functions $f_i(\beta)$
with the five-dimensional spherical harmonics
$\Psi_{v\alpha{}LM}(\gamma,\vartheta)$.

This method is especially suitable for application to transitional and
deformed nuclei, since the radial basis functions can be chosen to
match the particular radial potential at hand.  It provides vastly
more rapid convergence in the presence of significant $\beta$
deformation than is obtained with conventional methods based on
diagonalization in an oscillator basis (see
Ref.~\cite{rowe2005:collective-algebraic}).  Also, it can be applied
to potentials for which the oscillator basis methods are simply
inapplicable.  For the $\xfive$ problem, the wave functions must
vanish for $\beta\<>1$.  This boundary condition cannot be satisfied
in a finite basis of oscillator eigenfunctions.  Instead, in the
present work suitable basis functions are defined as
$f_i(\beta)\<\propto\beta^{-3/2}J_\nu(x_{\nu,i}\beta)$ for $\beta\<\leq1$
and zero elsewhere, where $x_{\nu,i}$ is the $i$th zero of $J_\nu$.
The calculation of Hamiltonian matrix elements with respect to this
radial basis is described further in the Appendix.  The choice
$\nu\<=3/2$ makes these $f_i(\beta)$ the exact radial wave functions
for the seniority zero states in the $\efive$ limit
($a\<=0$)~\cite{iachello2000:e5}.

Rowe \textit{et al.}~\cite{rowe2004:spherical-harmonics} provide an
algorithm for the explicit construction of the five-dimensional
spherical harmonics, as sums of the form
$\Psi_{v\alpha{}LM}(\gamma,\vartheta)\<=\sum_{\substack{K=0\\\text{even}}}^L
F_{v\alpha{}LK}(\gamma)\phi^L_{MK}(\vartheta)$, where
$\phi^L_{MK}(\vartheta)\<\equiv[(2L+1)/(16\pi^2(1+\delta_K))]^{1/2}[D^L_{MK}(\vartheta)+(-)^LD^L_{M-K}(\vartheta)]^*$.
The spherical harmonics are seniority eigenstates, labeled by the
seniority quantum number $v$, a multiplicity index $\alpha$, and the
angular momentum quantum numbers $L$ and $M$.  They are the exact
angular wave functions in the $\gamma$-soft ($a\<=0$) limit of the
present problem.  In the Hamiltonian~(\ref{eqnH}), the angular kinetic
energy operator (the quantity in parentheses) is simply the negative
of the seniority operator
$\hat\Lambda$~\cite{wilets1956:oscillations}.  Thus, calculation of
its matrix elements between spherical harmonics is
trivial~(\ref{eqngammakeme}).  The matrix elements of an arbitrary
spherical tensor function of $\gamma$ and the Euler angles can be
obtained through a series of straightforward integrations, as detailed
further in the Appendix.

High-seniority spherical harmonics are needed for the construction of
highly $\gamma$-localized wave functions.  Thus, in general,
diagonalization for larger $\gamma$-stiffnesses requires larger
angular bases.  For the range of $\gamma$ stiffnesses considered in
the present work ($0\<\leq a\<\leq1000$), a product basis constructed
from the first $\rtrim\sim5$ radial functions and the $15$ to $20$
lowest seniority spherical harmonics suffices to provide convergence of the
calculated observables for low-lying states.

\section{Results}
\label{secresults}
\begin{figure*}
\begin{center}
\includegraphics[width=0.7\hsize]{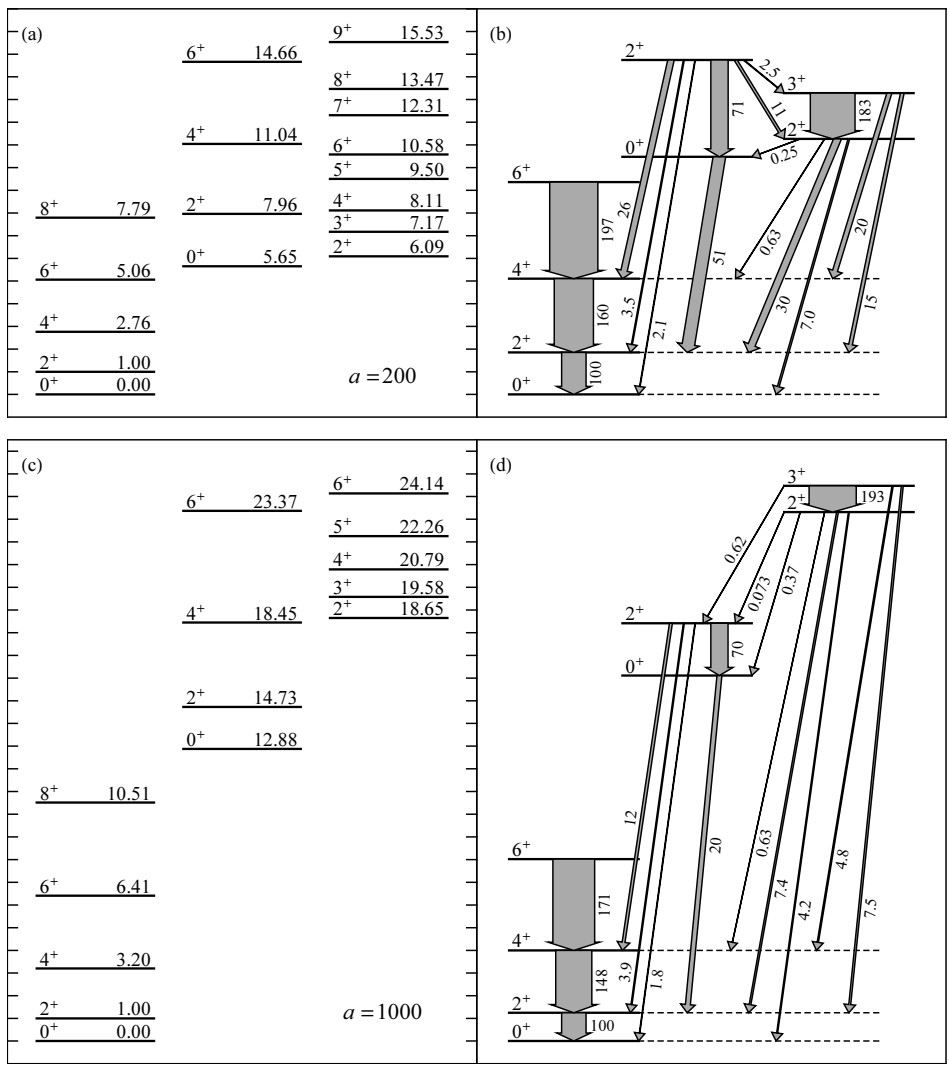}
\end{center}
\vspace{-12pt}
\caption
{ Level schemes for $a\<= 200$~(top) and $a\<= 1000$~(bottom), from
the exact numerical solution for the $\xfive$ Hamiltonian.  Excitation energies
of the lowest members of the ground, $\beta$, and $\gamma$ bands,
normalized to $E(2^+_1)$, are shown at left.  Electric quadrupole transition
strengths, normalized to $B(E2;2^+_1\rightarrow0^+_1)\<=100$, are
shown at right. }
\label{figa200a1000}
\end{figure*}
\begin{figure*}
\begin{center}
\includegraphics[width=0.8\hsize]{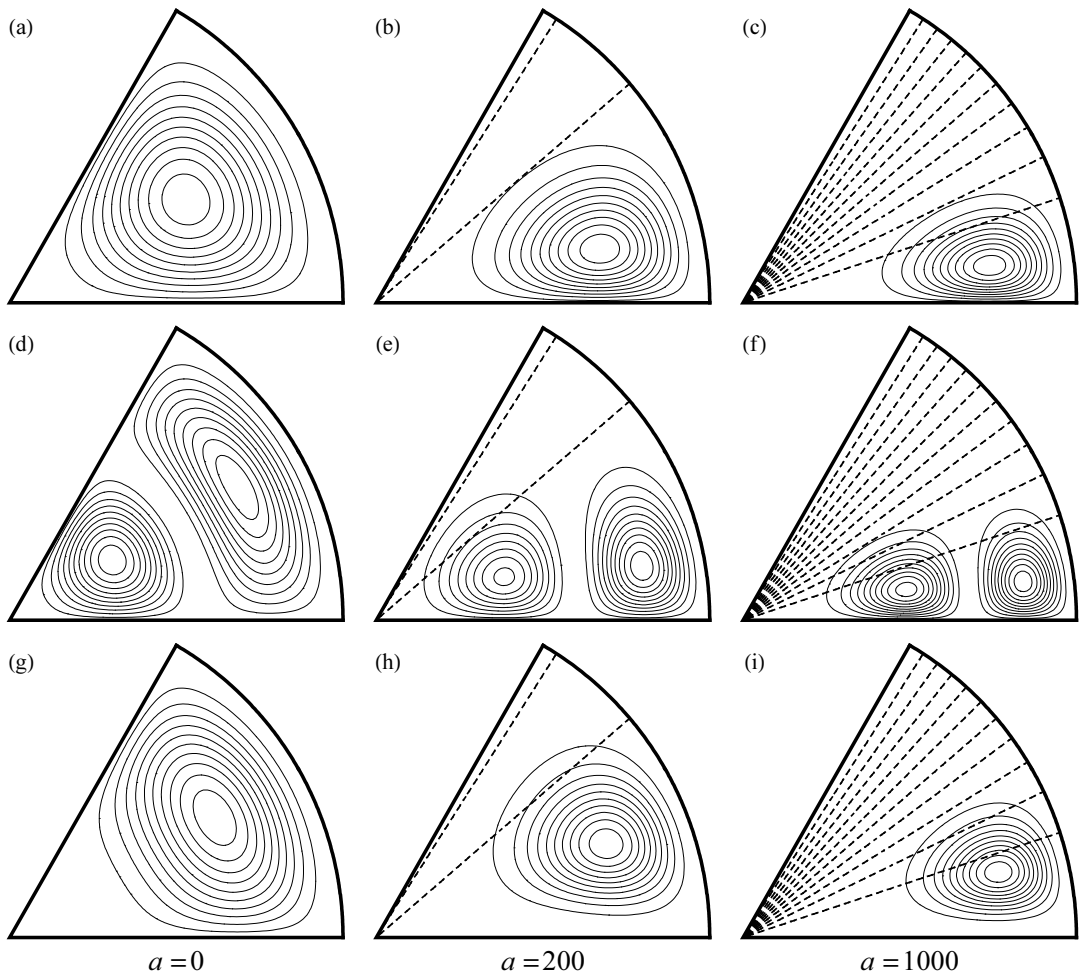}
\end{center}
\vspace{-12pt}
\caption
{Contour plots of the probability distributions with respect to
$\beta$ and $\gamma$, from the exact numerical solution for the
$\xfive$ Hamiltonian, for (top)~the ground state ($0^+_1$),
(middle)~the $\beta$ band head ($0^+_\beta$), and (bottom)~the
$\gamma$ band head ($2^+_\gamma$).  These are shown for $\gamma$
stiffness parameter values $a\<=0$~(left), $a\<=200$~(middle), and
$a\<=1000$~(right).  Contours of the potential
[$V_\gamma(\gamma)\<=a\gamma^2$] are also shown (dashed curves).  For
the $\gamma$-soft case $a\<=0$, where the bands are not well defined,
the $0^+_1$ ($\xi\<=1$, $v\<=0$), $0^+_2$ ($\xi\<=2$, $v\<=0$), and
$2^+_2$ ($\xi\<=1$, $v\<=2$) states, respectively, are shown.  Plots
are in the standard polar form, with $\beta$ as the radial
coordinate and $\gamma$ as the angular coordinate.  }
\label{figbetagammadistribpolar}
\end{figure*}
\begin{figure*}
\begin{center}
\includegraphics[width=0.7\hsize]{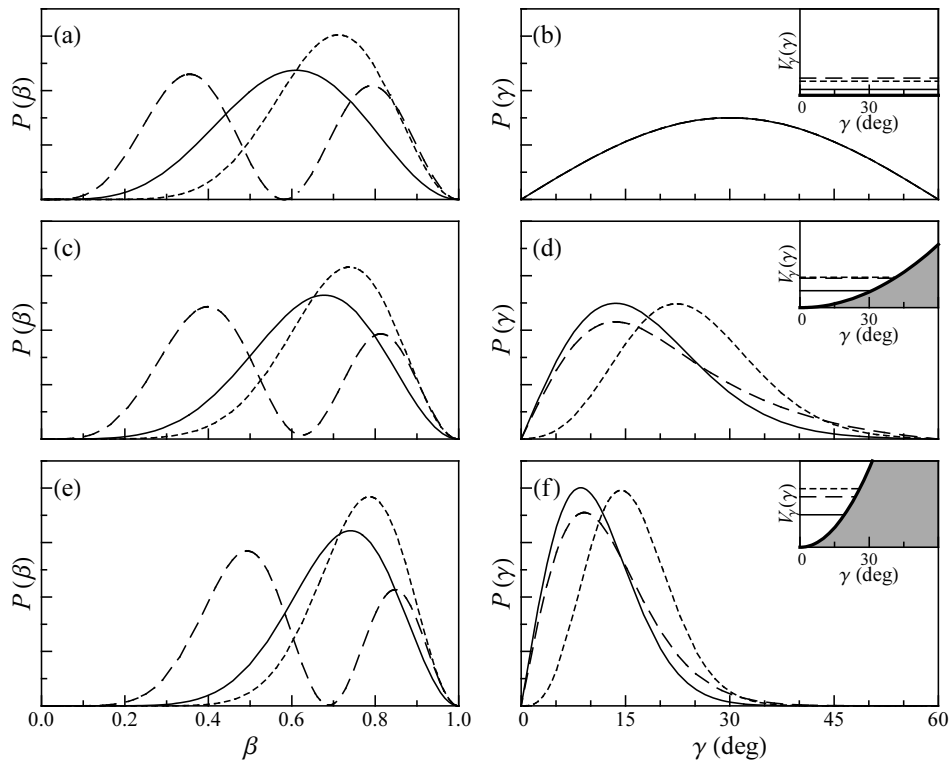}
\end{center}
\vspace{-12pt}
\caption
{Probability distributions with respect to $\beta$~(left) and
$\gamma$~(right) for the ground state (solid), the $\beta$ band head
(dashed), and the $\gamma$ band head (dotted), shown for $\gamma$
stiffness parameter values (top)~$a\<=0$, (middle)~$a\<=200$, and
(bottom)~$a\<=1000$.  The $\gamma$-confining potential $V_\gamma$ and
the level eigenvalues are shown in the insets. In panel~(b), as in
Fig.~\ref{figbetagammadistribpolar}, the $0^+_1$ ($\xi\<=1$, $v\<=0$),
$0^+_2$ ($\xi\<=2$, $v\<=0$), and $2^+_2$ ($\xi\<=1$, $v\<=2$) states
are shown.  These states have identical $\gamma$ distributions,
$P(\gamma)\propto\sin3\gamma$.  }
\label{figbetagammadistribcombo}
\end{figure*}
\begin{figure}
\begin{center}
\includegraphics[width=\hsize]{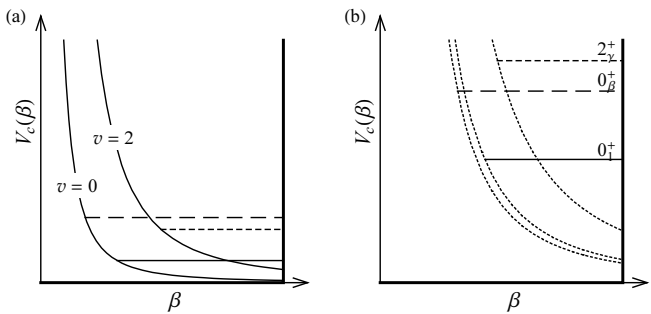}
\end{center}
\vspace{-12pt}
\caption
{(a)~Centrifugal potentials $V_c(\beta)\<=(v+1)(v+2)/\beta^2$ for the
radial Schr\"odinger equation in the $\gamma$-soft limit, for quantum
numbers $v\<=0$ and $2$.  (b)~Effective centrifugal potentials
$V_c(\beta)\<=\langle(v+1)(v+2)\rangle/\beta^2$, schematically
indicating the strength of the centrifugal effect for the $\gamma$-stabilized
case $a\<=1000$.  The level energies and centrifugal potentials
are shown for the same states as in
Figs.~\ref{figbetagammadistribpolar}
and~\ref{figbetagammadistribcombo}, namely, the ground state
(solid), the $\beta$ band head (dashed), and the $\gamma$ band head
(dotted).  
  }
\label{figcentrif}
\end{figure}
\begin{figure}
\begin{center}
\includegraphics[height=3.5in]{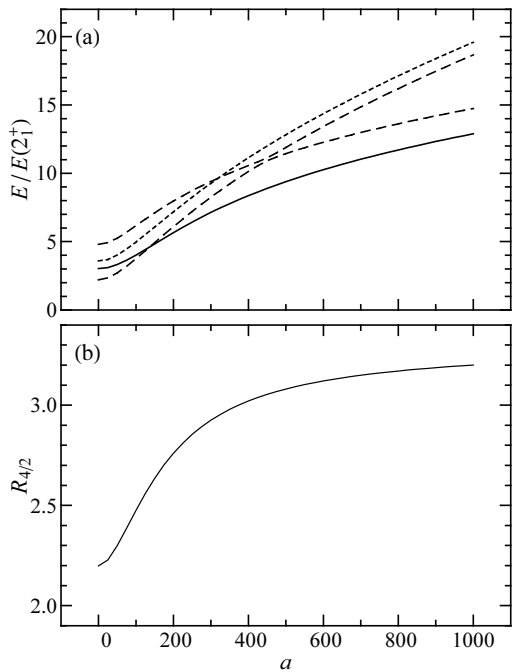}
\end{center}
\vspace{-12pt}
\caption
{Dependence of energy observables upon the $\gamma$ stiffness parameter $a$,
for $0\<\leq a\<\leq1000$.  (a)~Excitation energies of the $0^+_2$ (solid),
$2^+_2$ and $2^+_3$ (dashed), and $3^+_1$ (dotted) levels, normalized
to $E(2^+_1)$.  For large $a$, these states become 
the lowest spin members of the $\beta$ and $\gamma$ bands. 
(b)~Energy ratio $R_{4/2}\<\equiv E(4^+_1)/E(2^+_1)$.
}
\label{figevolne}
\end{figure}
\begin{figure}
\begin{center}
\includegraphics[height=3.5in]{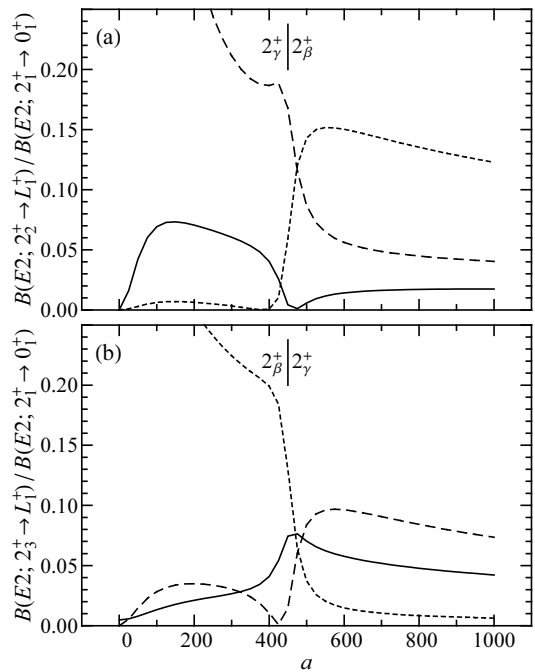}
\end{center}
\vspace{-12pt}
\caption
{Dependence of $B(E2)$ observables upon the $\gamma$ stiffness parameter $a$,
for $0\<\leq a\<\leq1000$.  (a)~Branches from the $2^+_2$ level to the
yrast $0^+$ (solid),
$2^+$ (dashed), and $4^+$ (dotted) levels, normalized
to $BE(2^+_1\rightarrow0^+_1)$.  Immediately below the avoided crossing at
$a\<\approx450$ this is the $2^+_\gamma$ level, while immediately
above it is the $2^+_\beta$ level (as indicated).
(b)~Branches from the $2^+_3$ level to the
yrast $0^+$ (solid),
$2^+$ (dashed), and $4^+$ (dotted) levels, normalized
to $BE(2^+_1\rightarrow0^+_1)$.  Immediately below the avoided crossing at
$a\<\approx450$ this is the $2^+_\beta$ level, while immediately
above it is the $2^+_\gamma$ level (as indicated).
}
\label{figevolnbe2}
\end{figure}
\begin{figure}
\begin{center}
\includegraphics[height=3.5in]{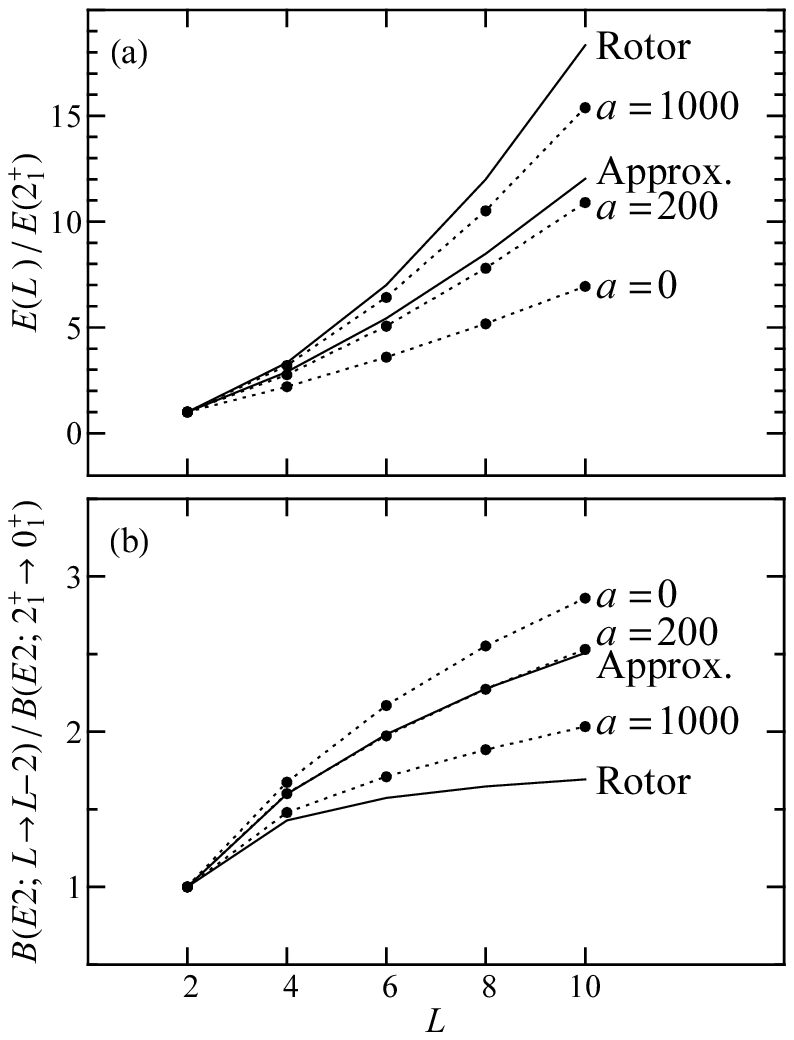}
\end{center}
\vspace{-12pt}
\caption
{Yrast band (a)~energies and (b)~$B(E2)$ strengths as functions of
angular momentum, for $a\<=0$, $200$, and $1000$ (dashed).  The values
obtained under the approximate separation of variables
(Sec.~\ref{secapprox}) are indicated for reference, as are the rigid
rotor values (solid). }
\label{figyrast}
\end{figure}

Results obtained from numerical diagonalization of the $\xfive$
Hamiltonian, for $a\<=200$ and $a\<=1000$, are shown in
Fig.~\ref{figa200a1000}.  Level energies and $E2$ transition strengths
for the lowest-lying levels are indicated.  More detailed tabulations
of energy and $B(E2)$ observables, calculated for $0\<\leq
a\<\leq1000$, are provided through the Electronic Physics Auxiliary
Publication Service (EPAPS)~\cite{axialsep-epaps}.  For the
transitional rare earth nuclei (with $N\<\approx90$) the $\gamma$ band
head energy is $\rtrim\sim6$ to $8$ times the $2^+_1$ energy,
consistent with a $\gamma$ stiffness $a\<\approx200$ to $300$.  This
may therefore be considered the most ``realistic'' range of values for
the $\gamma$ stiffness parameter, of the greatest phenomenological
interest.

First let us note some basic properties of the solution
[Fig.~\ref{figa200a1000}].  The levels are organized into
clearly-defined bands, characterized by strong intraband transition
strengths.  The lowest energy bands have the spin contents associated
with a rotational ground state band, $\beta$ band, and $\gamma$ band.  From a
comparison of the $a\<=200$ and $a\<=1000$ cases shown in
Fig.~\ref{figa200a1000}, it is apparent that such properties as the
band head excitation energies, energy spacings within the bands, and
transition strengths are strongly dependent on the $\gamma$ stiffness.
This is true for both the $\beta$ and $\gamma$ bands.  While it is
natural to expect the $\gamma$ band energy to increase with $\gamma$
stiffness, it is seen from Fig.~\ref{figa200a1000} that the $\beta$
band head energy also increases with $\gamma$ stiffness by a
comparable amount.  For larger $\gamma$ stiffness
[Fig.~\ref{figa200a1000}(c)], the energy spacings within the bands
tend towards rigid rotor [$L(L+1)$] spacings.

Before considering the spectroscopic observables in greater detail,
let us first examine the underlying features of the wave functions
$\varphi(\beta,\gamma,\vartheta)$.  In the approximate treatments of
the problem~\cite{iachello2001:x5}, the wave functions were separable
into products of $\beta$, $\gamma$, and Euler angle functions, which
were considered separately.  This is no longer possible for the full
solutions, but we can still consider the \textit{probability
distributions} with respect to one or more of these coordinates,
constructed by integration over all remaining coordinates (see
Appendix).  The integration metric appropriate to the Bohr coordinates
is $\beta^4d\beta
\lvert\sin 3\gamma \rvert d\gamma d\vartheta$~\cite{bohr1998:v2,eisenberg1987:v1}. Contour plots of the probability distribution with respect to $\beta$
and $\gamma$
are shown in Fig.~\ref{figbetagammadistribpolar}, for the ground,
$\beta$, and $\gamma$ band head states.  The probability distributions
with respect to $\beta$ or $\gamma$ individually,
\textit{i.e.}, integrated over the other coordinate, are shown for these
same states in
Fig.~\ref{figbetagammadistribcombo}.

The $\gamma$ dependences of the probability distributions of the
various states [Figs.~\ref{figbetagammadistribpolar}
and~\ref{figbetagammadistribcombo} (right)] are modulated by the
$\lvert\sin 3\gamma \rvert$ factor in the integration metric, which
causes the probability density to \textit{vanish} at $\gamma\<=0$ and
$\pi/3$.  In the $\gamma$-soft limit, the probability distributions
for several of the lowest-lying states are simply
$P(\gamma)\propto\sin3\gamma$ [Fig.~\ref{figbetagammadistribcombo}(b)]
and thus peaked at $\gamma\<=\pi/6$, or $30^\circ$.  For the $\gamma$-stabilized
cases [Fig.~\ref{figbetagammadistribcombo}(d,f)], the $\gamma$ distributions for the $0^+_1$ and
$0^+_\beta$ states are similar to each other, while that for the $2^+_\gamma$ state
is peaked at nearly twice the $\gamma$ value.  The 
probability distributions are compressed closer to $\gamma\<=0$ with increasing
Hamiltonian $\gamma$-stiffness, but even for the stiffest case
considered ($a\<=1000$) the value of
$\langle\gamma\rangle$ is $\rtrim\sim11^\circ$ for the ground state
and $\rtrim\sim16^\circ$ for the $\gamma$ band head.

Thus, for realistic $\gamma$ stiffnesses, the wave functions exhibit
considerable ``dynamical'' $\gamma$ softness.  This is perhaps
contrary to the common conception that nuclei with well-defined
$\gamma$-bands, like those in Fig.~\ref{figa200a1000}, are ``axially
symmetric'' and have $\gamma\<\approx0$.  (Recent empirical estimates
of the effective $\gamma$ deformation of transitional and deformed
rare earth nuclei~\cite{werner2005:triax-invariant} yield comparably
large dynamical $\gamma$ softness.)

The $\beta$ dependences of the probability distributions of the
various states [Figs.~\ref{figbetagammadistribpolar}
and~\ref{figbetagammadistribcombo} (left)] are seen to be strongly
influenced by $\beta$-$\gamma$ interaction, migrating towards larger
$\beta$ as the $\gamma$ stiffness increases.  This evolution can be
understood in a qualitative fashion in terms of the five-dimensional
analogue of the centrifugal effect.  In the $\gamma$-soft limit, where
the problem is separable, the $\beta$ wave function is governed by a
radial Schr\"odinger equation containing a term
$V_c(\beta)\<=(v+1)(v+2)/\beta^2$~\cite{wilets1956:oscillations,rakavy1957:gsoft},
analogous to the centrifugal potential in the three-dimensional
central force problem, as shown in Fig.~\ref{figcentrif}(a).  The
strength of this term depends upon the seniority quantum number $v$,
alternatively denoted $\tau$ in Ref.~\cite{iachello2000:e5}.  The
$1/\beta^2$ dependence energetically penalizes small $\beta$ values and
tends to ``push'' the wave function towards larger $\beta$.  But even
in the $\gamma$-stabilized situation, an analogue of the centrifugal
effect occurs, and its general features may be understood by observing
that the angular kinetic energy operator [the factor in parentheses in
the Hamiltonian~(\ref{eqnH})] multiplies a $\beta$-dependent factor
$1/\beta^2$.  The angular kinetic energy contains contributions from
both the rotational degrees of freedom and the $\gamma$ degree of
freedom.  As the $\gamma$ stiffness of the potential increases, the
confinement about $\gamma\<=0$ results in a larger $\gamma$
contribution to kinetic energy, even for the ground state, where this
is essentially the $\gamma$ vibrational zero point kinetic energy.
Thus, as the $\gamma$ stiffness increases, the influence of the
$1/\beta^2$ term in the Hamiltonian becomes larger, displacing the wave
functions towards larger $\beta$.  

An idea of the strength of the centrifugal effect in the
$\gamma$-stabilized cases can be obtained by numerically evaluating
the expectation value of the angular kinetic energy for any given
state.  The resulting ``effective'' centrifugal potential
$V_c(\beta)\<=\langle(v+1)(v+2)\rangle/\beta^2\<=\langle\hat\Lambda+2\rangle/\beta^2$
is shown in Fig.~\ref{figcentrif}(b).  (This potential is not of
calculational value, but it is useful in understanding the solution
properties.)  For the square well $\beta$ potential used in the
present model, the extent of the wave function in $\beta$ is limited
by the hard wall at $\beta\<=1$.  The centrifugal effect compresses
the wave function against the wall, so for large $\gamma$ stiffnesses
the wave function becomes localized with respect to $\beta$, just
within the wall.  Thus, even though the \textit{potential} is flat in
$\beta$, the $\beta$-$\gamma$ interaction inherent in the
\textit{kinetic energy} operator induces something akin to rigid
$\beta$ deformation.  

The $0^+_\beta$ state has a bimodal probability distribution with
respect to $\beta$ [Fig.~\ref{figbetagammadistribcombo} (left)].  In
the $\gamma$-soft limit [Fig.~\ref{figbetagammadistribcombo}(a)], an
exact zero in the probability distribution arises from the node in the
radial wave function, but for $a\<>0$ an exact zero is not expected.
The $0^+_\beta$ state also has a smaller mean value of $\beta$ than
the ground state.  The $0^+_\beta$ state is subject to a similar
effective centrifugal potential to that for the ground state
[Fig.~\ref{figcentrif}(b)], but it has more energy available in the
$\beta$ degree of freedom and is thus less strongly confined against
the wall at $\beta\<=1$.  The $2^+_\gamma$ state instead has a larger
mean $\beta$ than the ground state.  The angular kinetic energy for
the $\gamma$ band is about twice that for the ground state, in the
limit of harmonic $\gamma$ oscillations~\cite{iachello2001:x5}, and
hence so is the centrifugal effect [Fig.~\ref{figcentrif}(b)].  For
$a\<=200$ the mean $\beta$ values are
$\langle\beta\rangle_{0^+_1}\<\approx0.64$,
$\langle\beta\rangle_{0^+_\beta}\<\approx0.54$, and
$\langle\beta\rangle_{2^+_\gamma}\<\approx0.70$, while for $a\<=1000$
they are $\langle\beta\rangle_{0^+_1}\<\approx0.71$,
$\langle\beta\rangle_{0^+_\beta}\<\approx0.60$, and
$\langle\beta\rangle_{2^+_\gamma}\<\approx0.76$.

The dependences of a few basic observables upon the $\gamma$ stiffness
are shown in Figs.~\ref{figevolne}--\ref{figyrast}.
Naturally, the $\gamma$ band energy increases with $\gamma$ stiffness:
the evolution proceeds from the $\efive$ limit ($a\<=0$), through
$\gamma$ excitation energies appropriate to rare earth transitional
nuclei ($a\<\approx200$ to $300$), to very $\gamma$-stiff structure
[Fig.~\ref{figevolne}(a)].  But the $\beta$ excitation
energy increases with $\gamma$ stiffness as well, at about half the
rate at which the $\gamma$ excitation energy does.  An avoided crossing of
the $2^+_\beta$ and $2^+_\gamma$ states occurs for $a\<\approx450$.

The angular momentum dependence of energies within the yrast band
varies substantially with $\gamma$ stiffness [Fig.~\ref{figyrast}(a)].
For large $a$, it approaches the $L(L+1)$ dependence of the axially
symmetric rigid rotor.  In particular, the energy ratio
$R_{4/2}\<\equiv E(4^+_1)/E(2^+_1)$ ranges from $2.20$ in the $\efive$
limit to $\gtrsim3.2$ for large $a$ [Fig.~\ref{figevolne}(b)].  The
yrast band $B(E2)$ strengths [Fig.~\ref{figyrast}(b)] also vary with
stiffness, for large $a$ likewise approaching rigid rotor values.  The
observables are thus seen to clearly reflect the rigid $\beta$
deformation induced by the centrifugal effect at large $\gamma$ stiffness.

Further inspection of Fig.~\ref{figa200a1000} provides a more detailed
view of the spectroscopic properties obtained.  Strong interband $E2$
transitions are predicted, some comparable in strength to in-band
transitions.  A radical suppression of the spin-\textit{descending}
$\beta$ band to ground state band transitions (\textit{e.g.},
$2^+_\beta\<\rightarrow0^+_1$) relative to the spin-\textit{ascending}
transitions (\textit{e.g.}, $2^+_\beta\<\rightarrow4^+_1$) is one of
the most notable features.  (Only the branching of the $2^+_\beta$
state is shown in Fig.~\ref{figa200a1000}, but the branchings of the
higher-spin band members are provided through the
EPAPS~\cite{axialsep-epaps}.)  The strengths of the interband
transitions depend in detail upon the $\gamma$ stiffness, as shown in
Fig.~\ref{figevolnbe2}.  As the $\gamma$ stiffness increases, 
the $\beta$ to ground band and $\gamma$ to ground band transitions
become weaker overall.  They also tend closer to the rigid rotor Alaga rule
branching ratios~\cite{bohr1998:v2}.  Strong $\beta$-$\gamma$
interband transition strengths occur as well [Fig.~\ref{figa200a1000}], but
these vary greatly with $\gamma$ stiffness.

For all $\gamma$ stiffnesses, the energy spacing scale of the $\beta$
band is enlarged relative to that of the ground state band
[Figs.~\ref{figa200a1000} and~\ref{figevolne}(a)].  (In contrast, the
spacing scale of levels within the $\gamma$ band is similar to that of
the ground state band.)  The enlarged spacing scale may be understood
in terms of the $\beta$ probability distributions.  For a rigid rotor,
with fixed $\beta$, the spacing scale of levels within a band is
proportional to $1/\beta^2$~\cite{bohr1998:v2}.  As observed above,
the probability distribution of the $0^+_\beta$ state tends towards
smaller $\beta$ than that of the ground state, giving a larger average
$1/\beta^2$.  The ratio
$\langle\beta^{-2}\rangle_{0^+_\beta}/\langle\beta^{-2}\rangle_{0^+_1}$
has a maximum value of $\rtrim\sim2.1$ for $a\<\approx200$ and
decreases to $\rtrim\sim1.8$ for $a\<=1000$.  Similarly, the ratio of
spacing scales, $[E(2^+_\beta)-E(0^+_\beta)]/E(2^+_1)$, has a maximum
value of $\rtrim\sim2.3$ for $a\<\approx200$ and decreases to
$\rtrim\sim1.9$ for $a\<=1000$.

From this understanding of the underlying mechanism, it is seen that
enlarged energy spacing scale of the $\beta$ band is an artifact of
the rigid well wall in the $\xfive$
Hamiltonian~\cite{caprio2004:swell}.  As already noted, the extra
energy available to the $\beta$ excitation, relative to the ground
state, allows its wave function to expand ``inward'' against the
centrifugal potential [Fig.~\ref{figcentrif}(b)], but not ``outward''
against the rigid wall.  This produces the larger
$\langle\beta^{-2}\rangle$ and hence rotational energy scale for the
$\beta$ excitation.  A similar expanded spacing scale for the $\beta$
band is encountered in descriptions of transitional nuclei with the
interacting boson model~(IBM)~\cite{scholten1978:ibm-u5-su3} and the
geometric collective model~(GCM)~\cite{zhang1999:152sm-gcm}.  In these
cases the potential wall is no longer rigid but rather quartic
($\rtrim\propto\beta^4$), but the same basic mechanism may apply (see,
\textit{e.g.}, Fig.~15.8 of Ref.~\cite{caprio2003:diss}).

Another distinctive feature of the exact solution is the staggering of
energies within the $\gamma$ band, clearly visible for $a\<=200$
[Fig.~\ref{figa200a1000}(a)].  The level energies are clustered as
$2^+(3^+4^+)(5^+6^+)\ldots$, contrary in sense to the
$(2^+3^+)(4^+5^+)\ldots$ staggering of the rigid triaxial
rotor~\cite{davydov1958:arm-intro}.  The staggering is a remnant of
the $\grp{SO}(5)$ multiplet
structure~\cite{wilets1956:oscillations,rakavy1957:gsoft} present in
the $\gamma$-soft limit ($a\<=0$) and is an observable manifestation
of the considerable dynamical $\gamma$ softness seen in
Fig.~\ref{figbetagammadistribcombo}(d).  It disappears with increasing
$\gamma$ stiffness [Fig.~\ref{figa200a1000}(c)].  In general in a
geometric description, a weakly $\gamma$-confining potential yields
both dynamical $\gamma$-softness and a low $\gamma$ excitation energy,
while a strongly $\gamma$-confining potential yields $\gamma$
localization and a high $\gamma$ excitation energy.  Hence, the
presence of signatures of $\gamma$ softness (such as the $\gamma$ band
staggering) is closely correlated with low $\gamma$ band energy.  The
quantitative relationship between these signatures depends upon the
particular potential $V(\beta,\gamma)$ used, so reproduction of the
$\gamma$ band staggering may prove to be a valuable phenomenological
test.  Staggering of approximately the calculated magnitude and sense
is indeed found in the $\gamma$ bands of rare earth transitional
nuclei (\textit{e.g.}, Ref.~\cite[Fig.~25]{scholten1978:ibm-u5-su3} or
Ref.~\cite{duslingXXXX:staggering}).  Note that residual $\gamma$-soft
staggering occurs for transitional structure in the interacting boson
model (IBM)~\cite{iachello1987:ibm} as well, also as a remnant of
$\grp{SO}(5)$ multiplet structure, disappearing in the $\grpsuthree$
limit, which for infinite boson number corresponds to rigid rotor
structure.

The exact solution of the $\xfive$ Hamiltonian has been seen here to
provide valuable insight into the effects dominating $\beta$-soft
transitional structure.  However, this Hamiltonian has several
limitations as a realistic Hamiltonian for detailed phenomenological
analysis.

(1) The rotational kinetic energy term in the
Bohr Hamiltonian,
\(\rtrim\propto
\sum_\kappa \hat{L}_\kappa^{\prime2}/[4\beta^2\sin^2(\gamma -\frac{2}{3} \pi \kappa)]
\), is
constructed using the irrotational flow moments of inertia,
\(
\mathscr{J}_\kappa\<=4B\beta^2\sin^2(\gamma -\frac{2}{3} \pi \kappa)
\).  There
is extensive empirical evidence~\cite{bohr1998:v2} that the actual
moments of inertia are intermediate between the irrotational and rigid
body values.  This has only been established for the overall
normalization of the moments of inertia, but it also calls into
question the proper $\beta$ dependence of these moments of inertia and
therefore the proper $\beta$ dependence of the accompanying $\gamma$
vibrational kinetic energy term.  This $\beta$ dependence is of
central importance, since it generates the $\beta$-$\gamma$ coupling
just found to play such a major role in the solution properties.

(2) The hard wall of the $\xfive$ square well potential introduces
unrealistic features to the solution, as already discussed.  The
compression of the wave function against the well wall by the
centrifugal effect was noted above to induce something approximating
rigid $\beta$ deformation.  For a softer well wall, the wave function
is instead free to expand to larger $\beta$, so the compression should
be gentler.  The tendency of observables towards rigid rotor values
for large $\gamma$ stiffness and the enlarged energy spacing within
the $\beta$ band may both be attenuated.

(3) The restriction of $V(\beta,\gamma)$ to the form
$V_\beta(\beta)+V_\gamma(\gamma)$ was imposed purely for convenience
in the approximate separation of variables~\cite{iachello2001:x5}.
Coupling terms such as $\beta^3
\cos{3\gamma}$ 
arise naturally in the classical limit of the
IBM~\cite{ginocchio1980:ibm-classical,dieperink1980:ibm-classical} and
have been used extensively in earlier work with the geometric model
(\textit{e.g.}, Refs.~\cite{gneuss1971:gcm,eisenberg1987:v1}).  The
importance of the $\beta$-$\gamma$ coupling in $V(\beta,\gamma)$ must
be explored.

The numerical techniques of
Refs.~\cite{rowe2004:tractable-collective,rowe2004:spherical-harmonics,rowe2005:collective-algebraic}
are flexible and can easily be applied to a broad range of
Hamiltonians.  It should therefore be straightforward to investigate
many of the effects just discussed.  Exact numerical diagonalization
of such arbitrary Hamiltonians may also be useful for more abstract
studies of the geometric Hamiltonian.  In particular, it is well known
that the geometric model and the IBM produce similar spectra under
certain circumstances.  But much remains to be understood about the
extent to which IBM coherent state energy
surface~\cite{ginocchio1980:ibm-classical,dieperink1980:ibm-classical}
can be identified with a geometrical model potential energy surface,
and the appropriate kinetic energy operator to be used with this
potential energy
surface (\textit{e.g.}, Refs.~\cite{vanroosmalen1982:diss,rowe2004:ibm-critical-scaling,rowe2005:ibm-geometric,garciaramos2005:ibm-u5-o6-quartic}).

\section{Approximate separation}
\label{secapprox}
\begin{figure*}
\begin{center}
\includegraphics[width=0.7\hsize]{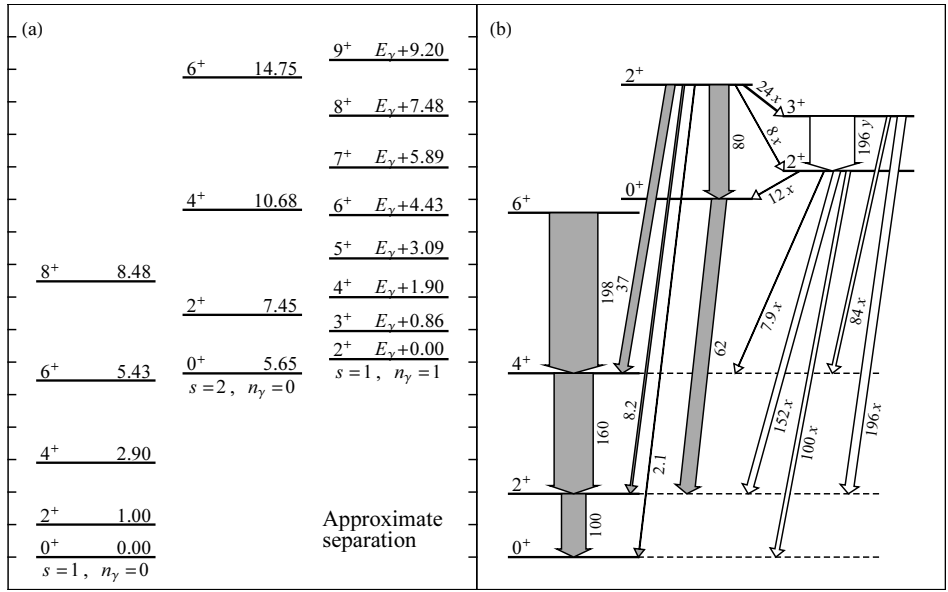}
\end{center}
\vspace{-12pt}
\caption
{ 
Level scheme for the approximate solution to the $\xfive$ Hamiltonian.
Excitation energies of the lowest members of the ground, $\beta$, and
$\gamma$ bands, normalized to $E(2^+_1)$, are shown at left.  Electric
quadrupole transition strengths, normalized to
$B(E2;2^+_1\rightarrow0^+_1)\<=100$, are shown at right.  The $\gamma$
band head energy $E_\gamma$ in part~(a) and the $\gamma$ band
transition strength normalizations $x$ and $y$ in part~(b) depend upon
the details of the $\gamma$ potential and are thus left unspecified.
For plotting purposes, the $\gamma$ band head position and arrow thickness
scales have been arbitrarily chosen to facilitate comparison with
Fig.~\ref{figa200a1000}~(top).  All observables involving the $\gamma$ band
are calculated according to the approximate separation as it
appears in Ref.~\cite{iachello2001:x5} rather than in
Ref.~\cite{bijker2003:x5-gamma}, which would yield values differing in
detail~\cite{endnote-bijkersep}.  
}
\label{figx5approx}
\end{figure*}

Since extensive prior work has been carried out using the approximate
separation of variables of Ref.~\cite{iachello2001:x5}, it is useful
to review this approximation, compare the results obtained under this
separation with the exact results, and establish more clearly the
reasons for the breakdown of the approximation.  Two simplifications
are required to obtain an approximate separation of variables for the
Hamiltonian~(\ref{eqnH}).  First, in the limit of small $\gamma$, the
sum $\sum_\kappa
\hat{L}_\kappa^{\prime2}/\sin^2(\gamma -2\pi \kappa/3)$ reduces to $4L(L+1)/3+K^2(1/\sin^2\gamma-4/3)$, where $L$ is the total
angular momentum quantum number and $K$ the intrinsic frame angular momentum
projection quantum number.  Second, in all terms involving $\gamma$,
the variable $\beta$ is replaced with a constant ``average'' value
$\beta_0$.  This yields a Hamiltonian $H= H_\beta+H_\gamma$
separable in the variables $\beta$ and $\gamma$,
\begin{widetext}
\begin{equation}
H=
\underbrace{ 
-\frac{1}{\beta^4}
\frac{\partial}{\partial \beta}
\beta^4  \frac{\partial}{\partial \beta}
+
\frac{L(L+1)}{3\beta^2}+V_\beta(\beta)
}_{H_\beta}
\,+\,
\underbrace{
{
\frac{1}{\beta_0^2}
\left[
-\frac{1}{\sin 3\gamma}
\frac{\partial}{\partial \gamma} \sin 3\gamma \frac{\partial}{\partial \gamma}
+\frac{K^2}{4}\left(\frac{1}{\sin^2\gamma}-\frac{4}{3}\right)
\right]
+V_\gamma(\gamma) } }_{H_\gamma}.
\label{eqnHapprox}
\end{equation}
\end{widetext}
The eigenvalues and eigenfunctions of $H_\beta$ are given by
$\varepsilon_\beta\<=x^2_{\nu,s}$ ($s\<=0,1,\ldots$) and
$f(\beta)\<\propto\beta^{-3/2}J_\nu(\varepsilon_\beta^{1/2}\beta)$
($0\<\leq\beta\<\leq1$), where $\nu\<=[L(L+1)/3+9/4]^{1/2}$ and
$x_{\nu,s}$ is the $s$th zero of $J_\nu$~\cite{iachello2001:x5}.
Under the small $\gamma$ approximation, the eigenproblem for
$H_\gamma$ reduces to that of the two-dimensional isotropic
oscillator, with quantum numbers $n_\gamma$ and $K$, and the
eigenvalues $\varepsilon_\gamma$ and eigenfunctions $\eta(\gamma)$ are
as described in Ref.~\cite{iachello2001:x5}.  The full eigenfunctions
of $H$ are products $f_{Ls}(\beta)\eta_{n_\gamma
K}(\gamma)\phi^L_{MK}(\vartheta)$ of a $\beta$ (``radial'') wave
function, a $\gamma$ wave function, and a rotational or Euler angle
($\vartheta$) wave function $\phi^L_{MK}(\vartheta)$, defined in terms
of $D$ functions as in Sec.~\ref{secsoln}.

Under these approximations, the $\xfive$ model yields levels arranged
in bands of good $K$ quantum number, but with level energies
[Fig.~\ref{figx5approx}(a)] and $E2$ transition strengths
[Fig.~\ref{figx5approx}(b)] which differ from those of the rigid rotor.  All
predictions for the ground state band and $\beta$ excitations are
independent of $V_\gamma$.  The spin dependence of level energies
within the yrast band is intermediate between those of the rigid rotor
and harmonic oscillator, with $R_{4/2}\<\equiv
E(4^+_1)/E(2^+_1)\<\approx2.90$.  The band arising from the first
$\beta$ excitation occurs at low energy
[$E(0^+_\beta)/E(2^+_1)\<\approx5.65$].  

Some of the characteristic spectroscopic features found in the
approximate solution (discussed in detail in, \textit{e.g.},
Refs.~\cite{iachello2001:x5,caprio2002:156dy-beta,bijker2003:x5-gamma})
are indeed encountered in the full solution (Sec.~\ref{secresults}).  The $\beta$ band
exhibits a substantially larger energy spacing scale than the yrast
band [$E(2^+_\beta)-E(0^+_\beta)\<\approx(1.80)E(2^+_1)$].  Strong
interband $E2$ transitions are predicted, as is the distinctive
branching pattern in which the spin-descending $\beta$ to
ground band transitions are suppressed.  

However, some of the essential features of the full solution are
missing from the approximate solution: the dependence of the $\beta$
band head energy upon $\gamma$ stiffness, the confinement of the wave
functions near $\beta\<=1$ due to the five-dimensional centrifugal
effect, and the consequent tendency of energy and transition strength
observables towards rotational values for large $\gamma$ stiffness.
Also, the observable signatures of $\gamma$-softness obtained in the
full solution for realistic $\gamma$ stiffnesses, such as the
staggering of energies in the $\gamma$ band, are absent under the
approximate separation of variables, which effectively enforces
$\gamma$ rigidity in the separation process.%
\begin{figure*}
\begin{center}
\includegraphics[width=0.6\hsize]{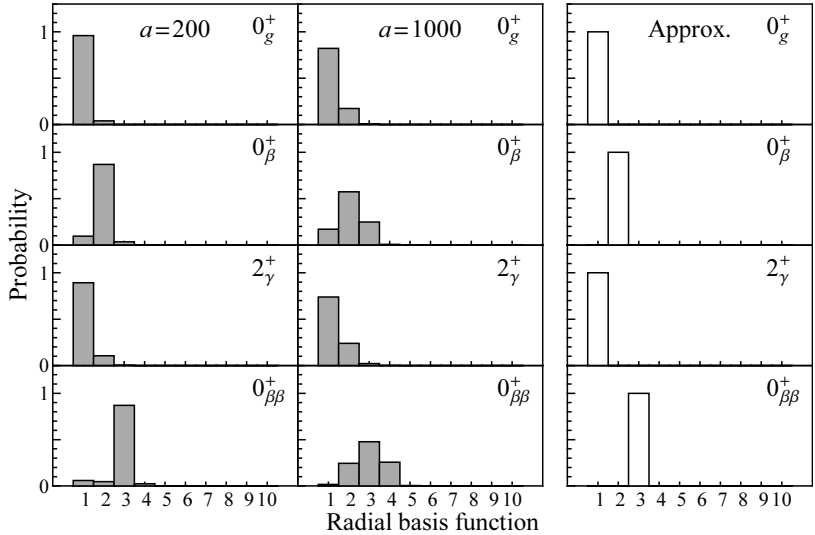}
\end{center}
\vspace{-12pt}
\caption
{Probability decompositions of the true $\xfive$ wave functions with
respect to the 
radial functions from the approximate solution, integrated over angular
coordinates.  Decompositions are shown for the ground, $\beta$,
$\gamma$, and $\beta\beta$ band heads, for $a\<=200$ (left) and
$a\<=1000$ (middle).  The relevant radial wave function under the
approximate separation is indicated (right) for comparison.
A basis of $L\<=0$ radial wave functions from the approximate solution
($\nu\<=3/2$) has been used for decomposition of the
$0^+$ states, while a basis of $L\<=2$ radial wave functions
($\nu\<=\sqrt{7}/2$) has been used for decomposition of the
$2^+_\gamma$ state.  
}
\label{figbetahist}
\end{figure*}

The basic reason for the failure of the approximate solution to
reproduce these features is clear from the analysis of
Sec.~\ref{secresults}.  The approximate Hamiltonian~(\ref{eqnHapprox})
retains only a portion of the centrifugal ($1/\beta^2$) contribution
to the full Hamiltonian~(\ref{eqnH}).  Through the term
$L(L+1)/(3\beta^2)$, the portion of the centrifugal effect arising
from \textit{rotational} kinetic energy is essentially retained.  But
the replacement of $1/\beta^2$ by $1/\beta_0^2$ in the remaining terms
suppresses the portion of the centrifugal effect arising from the
$\gamma$ kinetic energy.

For a more detailed understanding of the breakdown of the approximate
separation, we can reexamine the two approximations made in obtaining
Eqn.~(\ref{eqnHapprox}).

(1) The small angle approximation for $\gamma$
is in principle arbitrarily good for sufficiently large $\gamma$
stiffnesses.  However, as discussed in Sec.~\ref{secresults}
[Fig.~\ref{figbetagammadistribcombo} (right)], confinement to genuinely
small angles requires relatively large $\gamma$
stiffnesses ($a\<\gtrsim1000$).

(2) The validity of the other approximation, replacement of $\beta$
with $\beta_0$ in selected terms, depends upon one of two conditions
being met.  The approximation would be good if the wave function were
sharply localized in $\beta$, so indeed $\beta\<\approx\beta_0$.  This
condition may hold in some other contexts, such as small oscillations
for a rigid rotor~\cite{bohr1998:v2}, but it is not well
satisfied for the $\xfive$ problem, which was specifically constructed
to give a
\textit{lack} of $\beta$ localization.  Even without reference to the
exact solution, it is seen from the approximate radial wave functions
$f_{sL}(\beta)$ that the the probability distribution with respect to
$\beta$ is broad and differs significantly from state to state.  For
the approximate ground state $\langle 0^+_1
\vert \beta^{-2}
\vert 0^+_1\rangle\<\approx3.90$, while for the $\beta$ band head
$\langle 0^+_\beta \vert \beta^{-2} \vert
0^+_\beta\rangle\<\approx7.21$, nearly a factor of two larger.  Thus,
replacing $1/\beta^2$ by a ``rigid'' value $1/\beta_0^2$ is not
necessarily a small approximation.  Alternatively, the replacement of
$\beta$ with $\beta_0$ could still yield accurate results if the
overall strength of the approximated term were small.  This is seen
from Sec.~\ref{secresults} to occur when the kinetic energy in the
$\gamma$ degreee of freedom is small, and hence for nearly
$\gamma$-soft potentials.

The two approximations are thus valid at \textit{large} $\gamma$
stiffness and \textit{small} $\gamma$ stiffness, respectively.  If
these two regimes are to have any overlap, it must be at some intermediate
$\gamma$ stiffness.  Indeed, as noted above, there is considerable qualitative
agreement between the exact spectrum for $a\<=200$
[Fig.~\ref{figa200a1000} (top)] and the approximate spectrum
[Fig.~\ref{figx5approx}].  However, the significance of this
resemblance should not be overstated, since more detailed calculations
reveal that it arises in part from a cancellation of errors introduced
by the two approximations.  The small angle approximation tends to
cause an
\textit{overcalculation} of the $\beta$ band head energy, while the
rigid $\beta$ approximation tends to cause an \textit{undercalculation}.

A probability decomposition of the true wave functions with respect to
the approximate radial wave functions $f_{Ls}$, summed over angular
wave functions, provides a direct measure of
the resemblance of the true and approximate eigenstates.  This
decomposition is shown in Fig.~\ref{figbetahist}~(left, center) for
several states, for $a\<=200$ and $1000$.  The relevant radial basis
state from the approximate solution is also indicated, for comparison
[Fig.~\ref{figbetahist}~(right)].  For intermediate $\gamma$
stiffness, the approximate solution indeed dominates the decomposition of the
true solution, \textit{e.g.}, with
a $96\%$ probability for the $s\<=1$ basis state in the
true $a\<=200$ ground state [Fig.~\ref{figbetahist}~(left)].  The
similarity in wave functions breaks down for larger $\gamma$ stiffness
[Fig.~\ref{figbetahist}~(center)].

In summary, the approximate separation of variables of
Ref.~\cite{iachello2001:x5} causes the contribution of the $\gamma$
kinetic energy to the five-dimensional centrifugal effect to be
suppressed.  This is always a significant contribution and is the
dominant portion for large $\gamma$ stiffness.  The approximate
solution for the $\xfive$ Hamiltonian qualitatively reproduces some
aspects of $\beta$-soft transitional structure, but only those which
are least strongly affected by the $\beta$-$\gamma$ interaction.  The
approximate results \textit{quantitatively} resemble the exact results only for
a small range of $\gamma$ stiffnesses, in the vicinity of $a\<=200$.
This is approximately the $\gamma$ stiffness of
phenomenological interest for description of the rare earth
($N\<\approx90$) transitional region.  Consequently, many of the basic
conclusions found in prior comparisons with experimental
data~\cite{casten2001:152sm-x5,bizzeti2002:104mo-x5,kruecken2002:150nd-rdm,caprio2002:156dy-beta,dewald2002:150nd-rdm,caprio2003:diss,hutter2003:104mo106mo-rdm,clark2003:x5-search,tonev2004:154gd-rdm,mccutchan2004:162yb-beta,mccutchan2005:166hf-beta}
remain largely unaffected.

\section{Conclusion}
\label{secconcl}

The numerical techniques of Rowe \textit{et
al.}~\cite{rowe2004:tractable-collective,rowe2004:spherical-harmonics,rowe2005:collective-algebraic}
provide a practicable approach to the exact diagonalization of the
$\xfive$ Hamiltonian.  The wave functions and spectroscopic properties
obtained thereby provide insight into the structural features expected
for $\beta$-soft, axially stabilized transitional nuclei.  The
properties of the solution are found to be dominated by the
$\beta$-$\gamma$ coupling induced by the kinetic energy operator,
which results in a significant five-dimensional centrifugal effect.
Most spectroscopic properties are strongly dependent upon the $\gamma$
stiffness.  The results also highlight the presence of substantial
dynamical $\gamma$ softness.  These basic qualitative features were
not apparent from the approximate solution.  The analysis in the
present work can readily be extended to Hamiltonians which provide a
more realistic treatment of transitional nuclei.

\begin{acknowledgments}
Discussions with F.~Iachello and N.~Pietralla are gratefully
acknowledged.  This work was supported by the US DOE under grant
DE-FG02-91ER-40608 and was carried out in part at the European
Centre for Theoretical Studies in Nuclear Physics and Related Areas
(ECT*).
\end{acknowledgments}

\appendix
\section{Matrix elements}
\label{appme}

In this appendix a summary is provided of the procedure for
calculating the necessary matrix elements of operators between radial
or angular basis functions.  The matrices of operators in the full
product basis are obtained as the outer products of the radial and
angular matrices.  The radial basis functions $f_i(\beta)$ are defined
inside the well ($0\<\leq\beta\<\leq1$) as
$f_i(\beta)\<=A_{\nu,i}\beta^{-3/2}J_\nu(x_{\nu,i}\beta)$ and are
vanishing outside the well, where $x_{\nu,i}$ is the $i$th zero of
$J_\nu$ and
$A_{\nu,i}=[-J_{\nu-1}(x_{\nu,i})J_{\nu+1}(x_{\nu,i})]^{-1/2}$.  For
fixed $\nu$ and for $i\<=1$, $2$, $\ldots$, these form an orthonormal
set of functions with respect to the metric $\beta^4d\beta$.  The
angular basis functions are constructed as sums of the form
$\Psi_{v\alpha{}LM}(\gamma,\vartheta)\<=\sum_{\substack{K=0\\\text{even}}}^L
F_{v\alpha{}LK}(\gamma)\phi^L_{MK}(\vartheta)$ according to the
procedure of Ref.~\cite{rowe2004:tractable-collective}.  These are
orthonormal with respect to the metric $\lvert\sin3\gamma\rvert
d\gamma d\vartheta$.  The $\phi^L_{MK}$ are symmetrized combinations
of $D$ functions,
$\phi^L_{MK}(\vartheta)\<\equiv[(2L+1)/(16\pi^2(1+\delta_K))]^{1/2}[D^L_{MK}(\vartheta)+(-)^LD^L_{M-K}(\vartheta)]^*$,
and are orthonormal under integration over the Euler angles, except
that $\phi^L_{MK}\<=0$ in the special case of $K\<=0$ and $L$ odd.

Matrix elements of an arbitrary function $g(\beta)$ between
radial basis functions are calculated by 
straightforward numerical integration, as
\begin{equation}
\langle
f_{i'}
\vert
g(\beta)
\vert
f_i
\rangle\<=
\int \beta^4 d\beta f_{i'}(\beta) g(\beta) f_{i}(\beta).
\end{equation}
The matrix elements of the radial kinetic
energy operator in Eqn.~(\ref{eqnH}) can be reexpressed in terms of
matrix elements of $\beta^{-2}$ by use of the Bessel equation, as
\begin{multline}
\langle
f_{i'}
\vert
\left(
\frac{1}{\beta^4}
\frac{\partial}{\partial \beta}
\beta^4  \frac{\partial}{\partial \beta}
\right)
\vert
f_i
\rangle
\\
=x_{\nu,i}^2\delta_{i'i}
-\left(\nu^2-9/4\right)
\langle
f_{i'}
\vert
\beta^{-2}
\vert
f_i
\rangle
.
\label{eqnbetakeme}
\end{multline}
Observe that for the $\xfive$ Hamiltonian the matrix elements of
$V_\beta$ vanish, since $V_\beta(\beta)\<=0$ inside the well.  The
presence of the square well radial potential thus enters the calculations
only implicitly, through the boundary condition it places on the
allowed basis functions, which in turn dictates their kinetic energy matrix
elements~(\ref{eqnbetakeme}).

The matrix elements of an arbitrary function $f(\gamma)$, such 
as the function $\gamma^2$ appearing in the angular potential, can
be calculated through a series of straightforward
integrations involving the coefficients $F_{v\alpha{}LK}(\gamma)$, 
as~\cite{rowe2004:tractable-collective}
\begin{multline}
\langle\Psi_{v^\prime \alpha^\prime L } \Vert f(\gamma) \Vert
\Psi_{v \alpha L }\rangle
=
\langle\Psi_{v^\prime \alpha^\prime LL } \vert f(\gamma) \vert \Psi_{v \alpha LL }\rangle
\\
=\sum_{\substack{K\\\text{even}}}
\int \lvert\sin 3 \gamma\rvert d \gamma F_{v^\prime \alpha^\prime L
K }^*(\gamma) f(\gamma)  F_{v
\alpha L K }(\gamma) 
.
\label{eqngammascalarme}
\end{multline}
Here the Wigner-Eckart normalization convention of
Rose~\cite{rose1957:am} has been used in the definition of the reduced
matrix element.  The integration can be restricted to
$0\<\leq\gamma\<\leq\pi/3$ by periodicity of the functions in
$\gamma$~\cite{rowe2004:tractable-collective}.  The
expression~(\ref{eqngammascalarme}) is readily extended, by
application of the Clebsch-Gordan series, to give the matrix element
of any spherical tensor operator
$f^\lambda_\mu(\gamma,\vartheta)$, provided it is expanded in terms of
$D$ functions as
$f^\lambda_\mu(\gamma,\vartheta)\<=\sum_{\substack{\kappa=0\\\text{even}}}^\lambda
f^\lambda_\kappa(\gamma)\phi^{\lambda\,*}_{\mu\kappa}(\vartheta)$.
The
matrix element is
\begin{widetext}
\begin{multline}
\langle\Psi_{v^\prime \alpha^\prime L^\prime } \Vert f^\lambda(\gamma,\vartheta) \Vert \Psi_{v \alpha L }\rangle
=\frac{1}{4\pi}
\left[\frac{(2L +1)(2\lambda+1)}{2L^\prime +1}\right]^{1/2}
\sum_{\substack{K^\prime ,\kappa,K \\\text{even}}}
\left[\frac{1+\delta_{K^\prime }}{(1+\delta_{K })(1+\delta_{\kappa})}\right]^{1/2}
\\\times
\left[ (L K \lambda\kappa | L^\prime K^\prime ) 
+ \begin{cases*}
(-)^\lambda(L K \lambda\bar\kappa | L^\prime  K^\prime ) & K \geq\kappa\\
(-)^{L }(L \bar K \lambda\kappa | L^\prime  K^\prime ) & K \leq\kappa
  \end{cases*}
\right]
\left[\int \lvert\sin 3 \gamma\rvert d \gamma F_{v^\prime \alpha^\prime L^\prime K^\prime }^*(\gamma) f^\lambda_\kappa(\gamma)  F_{v \alpha L K }(\gamma) \right]
.
\label{eqngammatensorme}
\end{multline}
\end{widetext}
For the leading order $E2$ transition
operator~\cite{bohr1998:v2},
\begin{multline}
\label{eqnme2intrinsic}
\mathfrak{M}(E2;\mu)
\propto
\beta\bigl[D^{2\,*}_{\mu0}\cos\gamma
\\
+\frac{1}{\sqrt{2}}(D^{2\,*}_{\mu2}+D^{2\,*}_{\mu-2})\sin\gamma\bigr]
,
\end{multline}
the expansion coefficients are
$f^2_0(\gamma)\<\propto(8\pi^2/5)^{1/2}\cos\gamma$ and
$f^2_2(\gamma)\<\propto(8\pi^2/5)^{1/2}\sin\gamma$.  
Transition strengths are
$B(E2;i\<\rightarrow
f)\<=\linebreak[0](2L_f+1)\linebreak[0]\lvert\langle\varphi_f\Vert
\mathfrak{M}(E2)\Vert\varphi_i\rangle\rvert^2\linebreak[0]/(2L_i+1)$.
The angular
kinetic energy operator is the negative of the seniority operator
$\hat{\Lambda}$~\cite{wilets1956:oscillations}.  Since the five
dimensional spherical harmonics are seniority eigenstates, the matrix
elements are simply
\begin{widetext}
\begin{equation}
\langle
\Psi_{v'\alpha{}'L'M'}
\vert
\Bigl(
\frac{1}{\sin 3\gamma} 
\frac{\partial}{\partial \gamma} \sin 3\gamma \frac{\partial}{\partial \gamma}
\\ - \frac{1}{4}
\sum_\kappa \frac{\hat{L}_\kappa^{\prime2}}{\sin^2(\gamma -
\frac{2}{3} \pi \kappa
)}
\Bigr)
\vert
\Psi_{v\alpha{}LM}
\rangle
=-v(v+3)\delta_{v'v}\delta_{\alpha'\alpha}\delta_{L'L}\delta_{M'M}.
\label{eqngammakeme}
\end{equation}
\end{widetext}

Finally, consider a wave function $\varphi(\beta,\gamma,\vartheta)$
decomposed in terms of the product basis functions as 
\(
\varphi(\beta,\gamma,\vartheta)\<=\sum_{i,k}a_{ik}f_i(\beta)\Psi_{kLM}(\gamma,\vartheta)
\), where $k$ is a shorthand for the indices $(v\alpha)$.  The
probability density with respect to $\beta$ and $\gamma$, integrated
over Euler angles, is 
\begin{multline}
P(\beta,\gamma)=\beta^4 \lvert \sin3\gamma\rvert
\sum_{\substack{K\\\text{even}}}
\sum_{\substack{i',k',i,k}}
a_{i'k'}a_{ik}
\\\times
f_{i'}(\beta)F_{k'LK}(\gamma)
f_{i}(\beta)F_{kLK}(\gamma)
.
\label{eqnbetagammaprob}
\end{multline}

\vfil


\input{axialsep.bbl}

\end{document}

%% file: axialsep.bbl
\providecommand{\ELSEVIER}{}
\newcommand{\identity}[1]{{#1}}